\documentstyle[12pt,aasms4]{article}
\lefthead{Osmer et al.}
\righthead{A Deep Multicolor Survey}

\begin{document}

\title{A DEEP MULTICOLOR SURVEY. IV. THE ELECTRONIC STELLAR CATALOG}

\author{Patrick S. Osmer and Julia D. Kennefick\altaffilmark{1}}
\affil{Astronomy Department,
The Ohio State University, 174 West 18th Avenue, Columbus, OH 43210;
posmer@astronomy.ohio-state.edu and julia@astro.ox.ac.uk}

\author{Patrick B. Hall\altaffilmark{2}}
\affil{Steward Observatory, University of Arizona, AZ 85721;
hall@astro.utoronto.ca}

\and

\author{Richard F. Green}
\affil{National Optical Astronomy
Observatories, P.O. Box 26732, Tucson, AZ 85726; rgreen@noao.edu}

\altaffiltext{1}{Now at the Department of Astrophysics, Nuclear
and Astrophysics Laboratory, Keble Road, Oxford OX1 3RH, England}
\altaffiltext{2}{Now at the Department of Astronomy, University of
Toronto, 60 St. George St., Toronto, Ontario, M5S 3H8 Canada}

\begin{abstract}
We make available in electronic form
the stellar catalog of 19,494 objects from
the Deep Multicolor Survey (DMS).  The DMS is based on CCD imaging
with the Mayall 4-m telescope
in {\it U,B,V,R$^\prime$,I75,I86} and covers 0.83 deg$^2$ in six fields
at high galactic latitude.  The survey reached $5\sigma$ limiting
magnitudes of 22.1 in $I86$ to 23.8 in $B$.  The catalog gives positions, 
magnitudes and error estimates, and classification codes
in the six filter bands for all the objects.
We present tables that summarize the spectroscopic results
for the 55 quasars, 44 compact
narrow emission-line galaxies, and 135 stars in the DMS that we have confirmed to date.  We also make available illustrations of all the
spectra.  The catalog and spectra can be obtained from the World Wide
Web at http://www.astronomy.ohio-state.edu/$\sim$posmer/DMS/.

\keywords{catalogs --- galaxies: compact ---  
quasars: general --- stars: general --- surveys}

\end{abstract}

\newpage

\section{Introduction}

The purpose of this paper is to make available to the 
community the stellar catalog from the Deep Multicolor
Survey (DMS) and the results we have obtained from follow-up spectroscopic observations of 234 objects in the
catalog.  The main objective of the DMS has been to search for fainter 
quasars at high redshift, $z > 3$, and at lower redshift, $z < 2.2$,
than had been covered before with multicolor images and a digital
detector.  In addition, the survey is useful for the study of
faint field galaxies and of faint stars at high galactic latitude.

The survey is described in three papers:
\cite{PAPER1} (Paper I), \cite{PAPER2} (Paper II),
and \cite{PAPER3} (Paper III).  Paper I describes the survey
fields and multicolor imaging data that were obtained with a 
2048$\times$2048 CCD camera on the Kitt Peak National Observatory
4-m Mayall Telescope.  It also covers the data reduction steps;
the detection, classification, and cataloging of objects; and the
photometric measurements and their error estimates.  Paper II covers
the different multicolor techniques we used to select quasar 
candidates in the survey, the initial spectroscopic observations
that we made with the Hydra multifiber
spectrograph, and the results we obtained for quasars and other
objects in the survey.  Paper III describes additional spectroscopic
observations we made of quasar candidates in the survey and summarizes
the results we have to date.

In this paper we summarize in \S2 the main properties of the survey.
We describe in \S3 the catalog of 19,494 stellar objects 
that we are making available in electronic form.  We review how 
quasar candidates were selected and the results of the follow-up
spectroscopic observations.  We also present in tabular form the
lists of quasars and compact narrow emission-line galaxies that
we have confirmed so far.  In addition, we give the results for
stars that were identified from the spectroscopic observations.  
Illustrations of the spectra of all the objects are available
electronically.  

One purpose in making these data available is to aid others carrying
out or planning new multicolor surveys.  For example, the data constitute
a valuable test or training set for new selection methods for different
kinds of objects.  Another purpose is to enable studies of the
objects in the catalog by all interested researchers.

\section{Properties of the Survey} 

The survey is based on CCD imaging with the 
Mayall 4-m telescope
of six fields covering 0.83 deg$^2$ of sky
at high galactic latitude.  The survey employed
six filters: $U,B,V,R',I75,I86$ and reached
5$\sigma$ limiting magnitudes of 22.1 to 23.8. We used
{\it FOCAS} (Valdes 1982) to detect and classify objects
in the CCD images.  

We found approximately 30,000 objects
in total, of which about half are galaxies.  For the 
purposes of our work we defined stellar objects as those
which were detected in three or more filters and classified
as stellar in half or more of those filters.  We adopted
this approach because we did not want to exclude active
galactic nuclei (AGNs) with barely detectable extensions
from their host galaxies.  We chose to accept the price
of greater contamination of the stellar catalog by galaxies
than a more strict requirement would produce.  However,
we modeled the selection effects to estimate the 
number of contaminating galaxies in the final catalog.

The result of our classification is a catalog of 19,494 stellar objects.
This is 9\% lower than the 21,375 objects quoted in Paper I due to a 
coding error which allowed into the final catalog objects not meeting
our stellar classification criteria.
This error did not affect Paper I's completeness estimates for stars
significantly, but did cause an overestimate of the galaxy contamination.
We now estimate the catalog's overall contamination rate to be 36.5\%
(7126 objects), with roughly half of the contaminants fainter than the
5~$\sigma$ limits, as before.
We emphasize that the contamination is expected to be $\sim$20\%
at the final catalog's 5~$\sigma$ limits.
Consult Paper I for more detailed information on these issues.

To illustrate some of the basic properties of the survey, we
show in Figure 1 the observed differential source counts
for the stellar objects in three different filters: {\it B, V,} and
{\it R$^\prime$}.  Note that the magnitudes at which the different
counts reach a peak (and decline rapidly toward fainter
magnitudes) agree well with the $5\sigma$ and 90\% completeness
magnitude of Table 3 in Paper I.  For reference, we show in Figure
2 the cumulative source counts for the same three filters.

A color-magnitude diagram offers a way to visualize other properties
of the catalog.  We show in Figure 3 a $V\ vs.\ B-V$ diagram
for the stellar objects in the catalog, which are mostly stars.  The
diagram shows that the bulk of the objects have $0.4 < B-V < 1.6$.
The variation in number density with color at constant {\it V}
magnitude is a result of both the stellar luminosity function and the
fact that for cooler stars, $B-V$ changes less as the temperature
decreases, producing an apparent piling up of stars toward
$B-V \approx 1.6$.  The diagonal cutoffs and lines at the
lower end of the diagram show 1) the different limits in the different
individual fields of the survey and 2) that the redder objects drop below
the limiting magnitude 
of the $B$ band at successively brighter $V$ magnitudes as $B-V$
increases.

\section{The Catalog}

Here we describe the contents and properties of the 
catalog.  We also present information about all the
quasars, compact narrow emission-line galaxies, and stars
we could identify from follow-up spectroscopy.  The
catalog itself is available through the DMS home page at 
http://www.astronomy.ohio-state.edu/$\sim$posmer/DMS/.
The stellar catalog contains the identification code, position
on the sky, X and Y coordinates on the image,
and the photometric magnitudes and error 
estimates for all six filters.  The positional
uncertainties are estimated to be 0\arcsec .5.  Note that the
identification code consists of the field identifier, according
to the nomenclature of Table 1 of Paper 1 plus a running
number for the object.  We have also provided for each object
the FOCAS classification codes (star, galaxy, diffuse, etc.) for each
image and filter.  The classification codes may be useful for different
approaches to separating stars and galaxies in the catalog. 

The photometric error estimates are important for many quantitative
uses of the catalog.  See \S\S4.4, 4.5, and 5.4 of
Paper I for a detailed description of our photometric procedure,
limiting magnitudes, and the error analysis.  In brief, we
verified that the IRAF PHOT task gave accurate error estimates
for our data, and we used PHOT to produce the individual error
estimates listed in the catalog.  Note that the error estimates incorporate
the cases for objects being fainter than the $3\sigma$ limits in some
of the individual frames (\S4.5 of Paper I).  
An error of 0.333 means that the magnitude value corresponds to
the $3\sigma$ limit for that object.  Also, a negative value of 0.333
means that the object should have been but was not detected in one
of the frames and thus the measurement is being flagged as potentially
subject to a non-statistical error.

\subsection{Candidate Selection}

As discussed in Paper II, quasar candidates for follow-up spectroscopy
were selected as outliers in six-dimensional magnitude space and
as outliers in different two-color diagrams designed to be most sensitive
to quasars with $z < 2.2$ and $z > 3$.  Consult Paper II for details
on the selection criteria and results of modeling the detection efficiency
as a function of magnitude and redshift.  Note also that there were
errors in the printing of some of the figures in Paper II, which were
reprinted correctly in a subsequent erratum (\cite{ERRATUM} 1996c).

\subsection{Quasars}

For reference and to provide a compilation in one place, 
we list in Table 1 information on all 55 of
the spectroscopically identified quasars we have found to date 
in the DMS.  The table contains both coordinate and catalog identification
numbers and the derived redshift along with the magnitudes and error
estimates in each filter.  Additional information on individual objects
is given in Papers II and III.

Table 1 is also available on-line via the DMS home page mentioned above.
It has the added feature that illustrations of the spectra are linked to the
object name and can be downloaded as desired.

\subsection{Compact Narrow-Emission Line Galaxies}

In Table 2 we provide information on compact narrow emission-line
galaxies (CNELGs) in identical format to that of Table 1. Note that
there are two entries for N2139-0400 because it falls in the overlap
region of fields 21e and 21w.  
Additional information
on individual objects is in Papers II and III.  As described
above for the quasars,
Table 2 and the spectra of the CNELGs are available from the 
DMS home page.

\subsection{Stars}

In Table 3 we summarize the information we have from the spectra of
stars that could be identified among the quasar candidates.  We list
catalog I.D. numbers, magnitudes and error estimates for the six
filters, and comments on the spectral features or type of the star.
As mentioned previously, we supply this information because it may 
be useful for researchers working on the properties of faint stars
at high galactic latitudes or working on plans for new surveys.  
However, we emphasize that the spectra were generally of rather
low signal-to-noise.  Therefore, users of this table should be aware
that our estimated spectral types are more uncertain than
is the norm in stellar research.  Finally, Table 3 and the spectra of
the stars are available on the DMS home page.

\acknowledgements

We would like to thank the KPNO mountain staff for their assistance
in observing and the KPNO TAC for their allocation of time for this
project.  We thank Jeannette Barnes and Frank Valdes for invaluable assistance with IRAF and FOCAS during the course of the survey.  We are grateful
to the referee, Gregory Aldering, for his suggestion about including the 
classification codes in the catalog.  This not only caused us to find
an error in our original computer code but should make the catalog
more useful to the community.
This work was supported in part by NSF Award AST-9529324.

%\begin{table}
%\tablenum{1}
%\dummytable\label{tbl-1}
%\end{table}
%\begin{table}
%\tablenum{2}
%\dummytable\label{tbl-2}
%\end{table}
%\begin{table}
%\tablenum{3}
%\dummytable\label{tbl-3}
%\end{table}

\clearpage

\figcaption[Osmer.fig1.eps]{The differential source counts
for the stellar objects in the DMS for three filters: $B,\ V,$ and
$R^\prime$.  The log of the number of objects in one magnitude bins 
is plotted {\it vs.} magnitude. \label{fig1}}

\figcaption[Osmer.fig2.eps]{The log of the cumulative source counts,
$(N < M)$, as a function of magnitude in $B,\ V,$ and $R^\prime$ for
the stellar objects in the DMS. \label{fig2}}

\figcaption[Osmer.fig3.eps]{The $V\ vs.\ (B-V)$ color-magnitude
diagram for the stellar objects in the DMS. \label{fig3}}

\clearpage

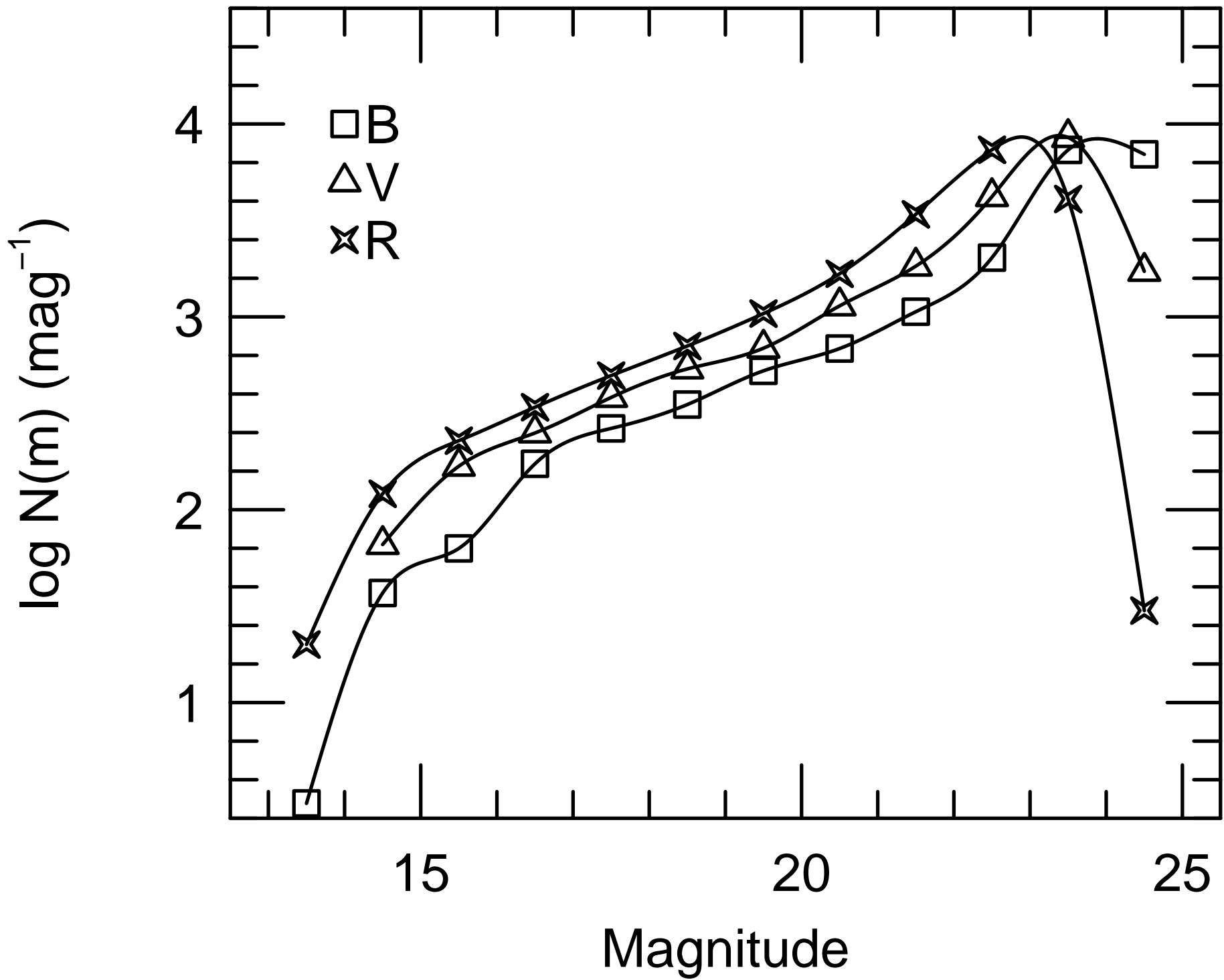
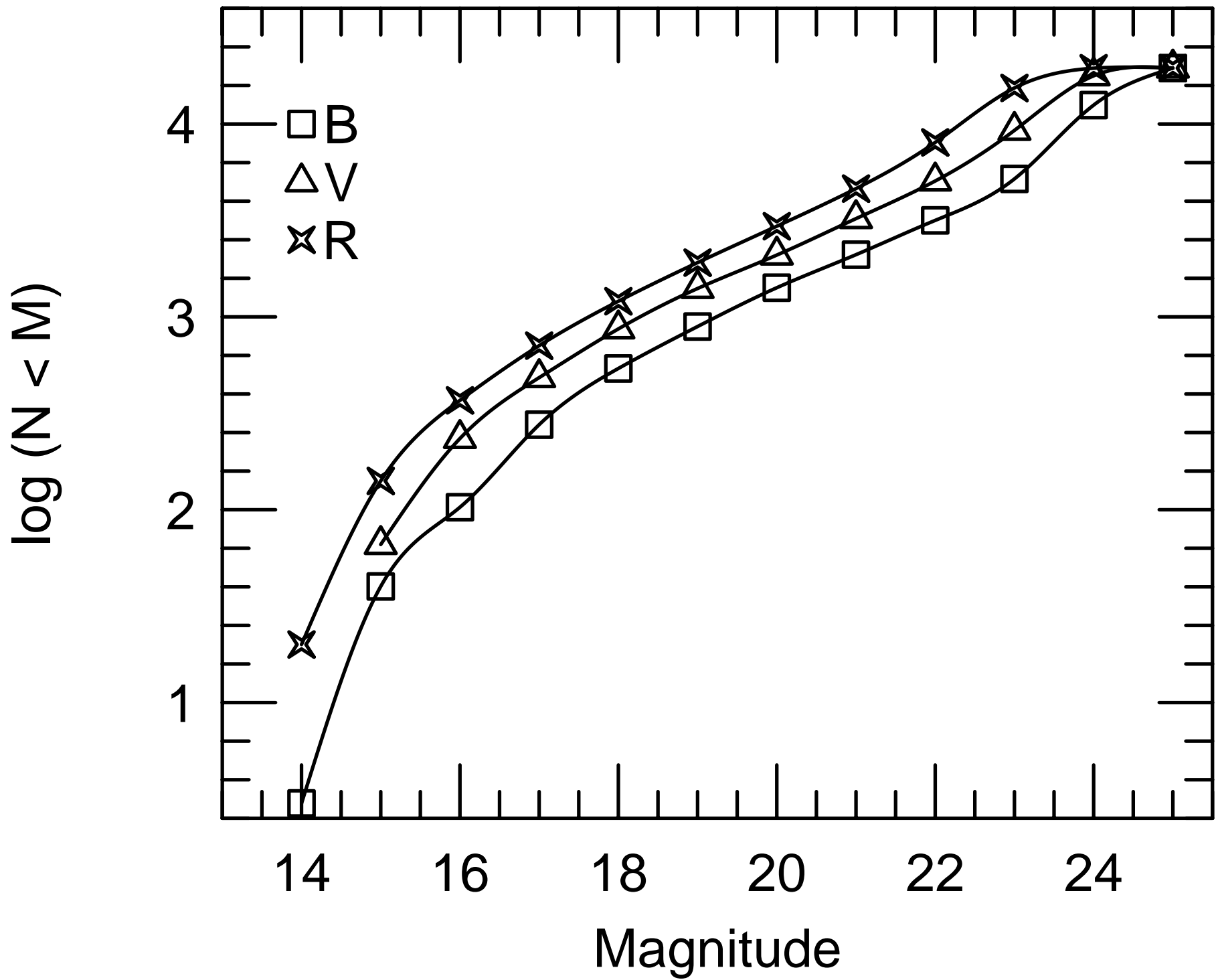
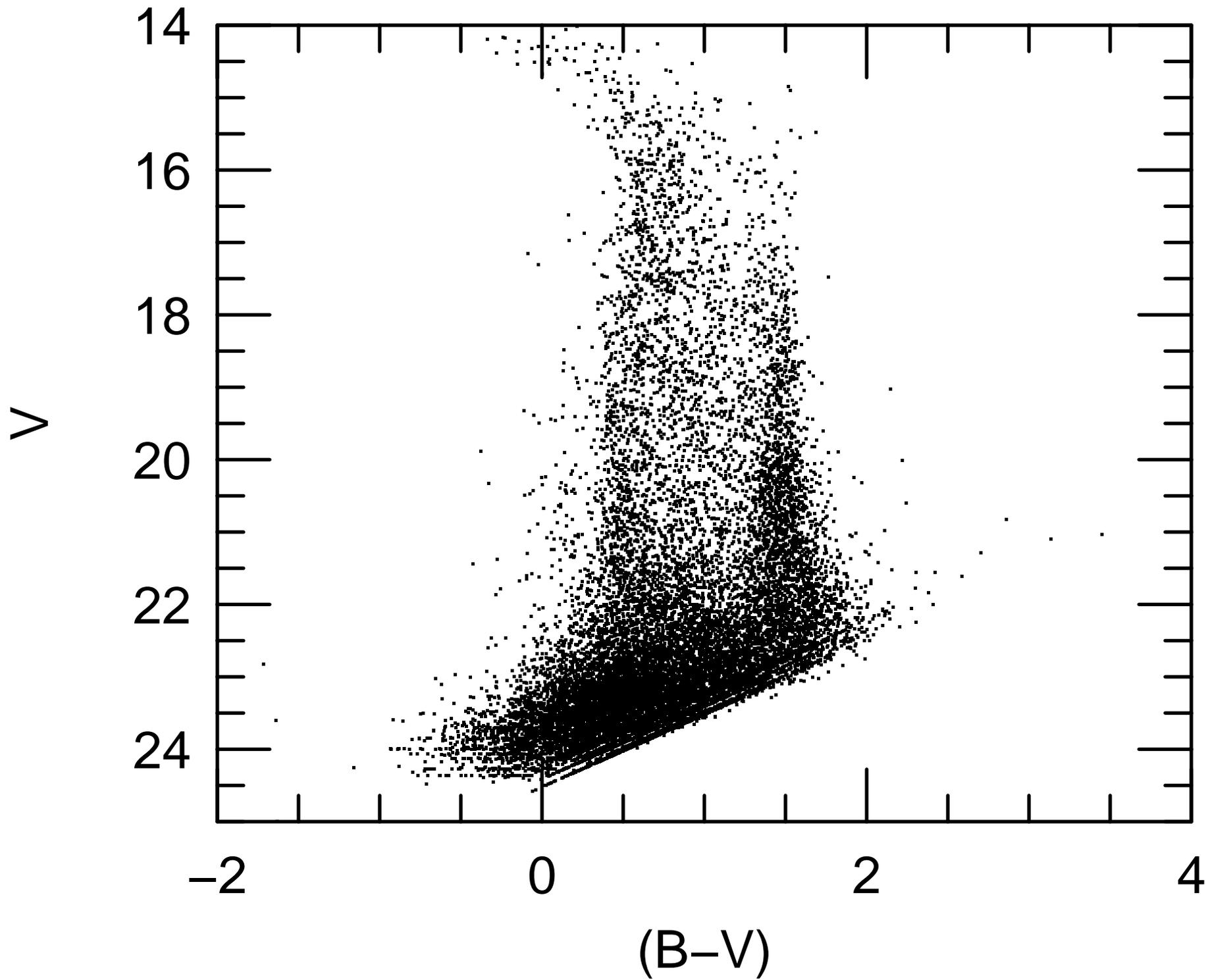
\end{document}